\documentclass[letterpaper,preprintnumbers,twocolumn,superscriptaddress,aps,nofootinbib,perprint]{revtex4}
\usepackage{CJK}\usepackage{amssymb}\usepackage[centertags]{amsmath}\usepackage{txfonts}\usepackage{epsfig}\usepackage{bm}\usepackage{color}\usepackage{graphicx,graphics}\usepackage{multirow}\usepackage{float}\usepackage{ulem}\usepackage{hyperref}\usepackage{setspace}\usepackage{slashed}
\usepackage{booktabs}\usepackage{geometry}\usepackage{diagbox}\usepackage{epstopdf}
\hypersetup{colorlinks=true, citecolor=blue, linkcolor=blue,filecolor=black,urlcolor=blue}
\allowdisplaybreaks[4]

\begin{document}

\begin{CJK*}{GBK}{song}

\title{Handedness Correlation from Quark Polarization}

\author{Weihua Yang}

\affiliation{College of Nuclear Equipment and Nuclear Engineering, Yantai University, Yantai, Shandong 264005, China}

\begin{abstract}
Jet handedness as a measure of quark and/or gluon polarizations has been proposed for nearly 30 years. It was demonstrated by measuring the correlation of jet handedness in the electron positron annihilation process.  Once parameters are determined, the method could be used to measure quark and/or gluon polarizations in other experiments. The reported data provided evidence for the jet handedness and handedness correlation. However,  the jet handedness correlation measured in the electron positron annihilation process from the opposite jets contradicts theoretical prediction by a sign. In order to explain this, we present a chromo-hydrogen-like model in this paper. According to calculations, both jet handedness and handedness correlation depend on not only the polarization of the fragmenting valence quark but also the polarization of the sea quark. It is the appearance of the sea quark polarization that can solve the contradiction.  In other words, measurements of jet handedness and handedness correlation can be used to determine the sea quark polarization. 
\end{abstract}

%\pacs{}

\maketitle

\section{Introduction}\label{sec-1}

Since the early days of collider physics, jet has been an important tool in study of quantum chromodynamics (QCD) which is a fundamental non-Abelian gauge theory of strong interactions of quarks and gluons. Jet plays an important role in the study of hadronization, color neutralization and quark confinement. Furthermore, almost 30 years ago, Efremov, Mankiewicz, T\"{o}rnqvist, Stratmann and Vogelsang proposed that quark and/or gluon polarizations (distributions) can be determined by measuring the jet handedness \cite{Efremov:1992pe,Tornqvist:1992ge,Stratmann:1992gu} which was introduced more than four decades \cite{Nachtmann:1977ek,Efremov:1978qy}.  Considering the chirality, jet handedness is defined by the probability difference of the right-handed and left-handed jets.  Measurement data provided evidence for the jet handedness and jet handedness correlations \cite{Efremov:1994uud,Efremov:1994va,Muller:1994tw,SLD:1994dmv,Efremov:1995fe,Efremov:1998zp}. However, experimental results of the handedness correlation from the opposite jets has a sign which contradicts the theoretical prediction.  This contradiction has no simple explanation. But it was considered as an evidence of the random chromo-magnetic field \cite{Efremov:1995fe}.

Shortly after  refs. \cite{Efremov:1992pe,Tornqvist:1992ge,Stratmann:1992gu}, Ryskin presented a simple model to explain the origin of handedness in the fragmentation process \cite{Ryskin:1993hu}. A quark-antiquark pair ($q, \bar q$) is produced in the color field induced by the isolated fast quark. Because of the opposite color charges and motion directions, the quark and the antiquark acquire transverse momenta in the same direction and therefore the jet handedness emerges.  This mechanism can also explain the handedness correlation in opposite jets by increasing the number of events with different handedness.  By considering the CP-violating effect of QCD vacuum, Efremov and Kharzeev proposed a heuristic method to illustrate the contradiction between experimental results and theoretical predictions \cite{Efremov:1995ff}. They constructed a two-particle fragmentation function describing a polarized quark fragmenting into a hadron pair in the background chromo-magnetic field of QCD vacuum and rewrote the handedness correlation with an auxiliary gluon condensate term, $\langle B^2\rangle$. The gluon condensate term dominates the sign of the handedness correlation. As authors pointed out this proposal could be a good indication of the presence of CP-violating effect in the fragmentation process, e.g, parity-violating fragmentation functions. The following discussions can be found in refs. \cite{Kang:2010qx,Yang:2019rrn,Yang:2019gdr,Yang:2020sos,Yang:2021xwn}.

Different from the previous discussions, in this paper, we provide an alternative method to explain the handedness correlation contradiction between experimental results and theoretical predictions. There is no need to introduce the CP-violating effect, by the way. We present a model in which the valence quark (Q) produced by the annihilation reaction and a quark (q) (antiquark ($\bar q$)) produced in the color field form a hydrogen-like system. %Calculation indicates the anti-parallel combination of particle spins has lower energy level and dominates the handedness correlation.
Based on this model, we recalculate the jet handedness and the handedness correlation and find that they depend on not only polarization of the fragmenting valence quark  (Q)  but also the polarization of the sea quark  (q) .  The polarization of the sea quark dominates the sign of the handedness correlation.

We note that the significance of the study of jet handedness and handedness correlation is not limited to explore quark and/or gluon polarizations. As mentioned in refs. \cite{Pisano:2015wnq,Metz:2016swz,Boer:2003ya}, jet handedness is considered to be strongly related to the di-hadron fragmentation functions  \cite{Konishi:1978yx,Bianconi:1999cd,Bianconi:1999uc,Bacchetta:2002ux,Boer:2003ya,Bacchetta:2003vn}. One of the reasons why one studies di-hadron fragmentation functions  is that they are universal and can be factorized in high energy reactions. By extracting from the two-jet events in the electron positron annihilation process \cite{Boer:2003ya,Bacchetta:2008wb,Courtoy:2012ry}, they can be used to study the nucleon structures,  especially for the transversity distribution function which reveals the transversely polarized quarks in a transversely polarized nucleon \cite{Bianconi:1999cd,Bacchetta:2008wb,Bacchetta:2004it}.
%Recently, conceptual design reports for the Circular Electron Positron Collider (CEPC) and the Electron Positron Collider: Future Circular Collider (FCC-ee) were released \cite{CEPCStudyGroup:2018ghi,FCC:2018evy}. Future electron positron colliders provide new significant opportunities to study the jet handedness, handedness correlations and di-hadron fragmentation functions and can help to uncover many hidden questions.

This paper is organized as follows. In sec. \ref{sec:handednessdef}, we present an introduction to jet handedness and handedness correlation based on previous discussions. The chromo-hydrogen-like model is given in sec. \ref{sec:model}. We note that this model is proposed to describe the first stage of the fragmentation process. We revisit the jet handedness and handedness correlation based on the  chromo-hydrogen-like model in sec. \ref{sec:handednessrevisit}. A brief summary will be given in sec. \ref{sec:summary}.

%First of all, we revisit the fragmentation process where a pair of spinless hadrons ($p_1, p_2$) is produced in a quark jet ($q$), see fig. \ref{fig:san},
%\begin{align}
% q\to p_1+p_2+X.
%\end{align}
%The asymmetric distribution of this process can be written as \cite{Efremov:1992pe}
%\begin{align}
 %\frac{d\sigma}{d\Omega} \sim (|f|^2+|g|^2)(1+\alpha \vec P \cdot \vec n),
%\end{align}
%where $f$ and $g$ denote the spin non-flip and spin flip amplitudes, respectively. $\vec P$ is the quark polarization and $\vec n$ is determined by momenta of produced hadrons, $\vec n=\vec p_1\times \vec p_2/|\vec p_1\times \bf p_2|$.
%$\alpha$ is the asymmetry parameter and defined as $\alpha= 2 \mathrm{Im} (f^* g)/ (|f|^2+|g|^2)$.

%\begin{figure}[t]
%  \centering
% \includegraphics[width=4cm]{san}\\
%  \caption{Illustration of directions of momenta in the fragmentation process.}\label{fig:san}
%\end{figure}

%In addition to the axial vector $\bf n$, we can introduce another pseudoscalar variable
%\begin{align}
% X= \bf n\cdot \bf j,
%\end{align}
%where $j$ is the unit vector in the jet direction defined by the total jet momentum of the thrust axis. Therefore, the jet is called right handed (left handed) if $X>0~(X<0)$. The handedness is defined as the difference of the probability of the right handed and left handed jets,
%\begin{align}
% H=\frac{N_R(X>0)-N_L(X<0)}{N}=\bar \alpha_W P,
%\end{align}
%where $N=N_R+N_L$ is the number of the total jets, $P$ is the quark polarization and $\alpha$ has been replaced by the average $\bar \alpha_W$ for some weighting procedure.

\section{Handedness and handedness correlation. I}  \label{sec:handednessdef}

In this section, we present a brief introduction to the jet handedness and handedness correlation based on the previous discussion in the electron positron annihilation process. Systematic discussions can be  found in references, e.g, \cite{Efremov:1992pe,Efremov:1995fe}.  We note that the jet handedness and the handedness correlation considered in this paper are longitudinal. Transverse jet handedness and handedness correlation(s) can be illustrated similarly. However, we do not show them due to lack of experiment data.

In order to introduce the jet handedness and the handedness correlation,  one first of all considers the process of a quark fragmenting into a pair of spinless hadrons and other unmeasured debris.
It is easily to construct the basic axial vector and pseudoscalar in terms of three-momenta of the produced hadrons ($\vec p_1, \vec p_2$) in the following forms,
\begin{align}
 &\vec n=\frac{\vec p_1\times \vec p_2}{|\vec p_1\times \vec p_2|}, \label{f:vecn} \\
 & X=\vec n \cdot \vec j, \label{f:scaX}
% & X_T=\vec n \cdot \vec e_T, \label{f:scaXT}
\end{align}
where $\vec j$ is the unit vector in the jet (momentum $k$) direction. It is defined as $\vec j= \vec k/|\vec{k}|$.  It can be seen that $\vec n$ is determined by hadron momenta ($\vec p_1, \vec p_2$) and the selection order according to definite criteria.

One can use $X$ to determine the jet. According to Eqs. (\ref{f:vecn})-(\ref{f:scaX}), a jet is called right-handed (left-handed) when $X>0~(X<0)$.
Therefore, the (longitudinal) handedness is defined as
\begin{align}
 H = \frac{N_R(X>0)-N_L(X<0)}{N} = \alpha P, \label{f:longH}
\end{align}
where subscripts $R, L$ indicate the right-handed and left-handed and  $N=N_R+N_L$. $P$ is the longitudinal polarization of the fragmenting valence quark ($Q$), $\alpha$ is a parameter which can determined in experiments. The physical meaning of $\alpha$ will be clear in the following context.

In practice, there are two kinds of selection rules to determine $p_1$ and $p_2$. They give different results to the jet handedness and handedness correlation. The first one (labeled by $Y$) is independent of electric charge, $p_1$ and $p_2$ can be chosen as $|p_1|>|p_2|$. In this case, $p_1$ may be the momentum of the leading particle in the fragmenting jet. The other criterion (labeled by $Q$) is dependent on electric charge. $p_1$ and $p_2$ can be chosen as the momenta of a positive (negative) charged and a negative (positive) charged particles, respectively.

For the parameter $\alpha$, we have
\begin{align}
 & \alpha_Y^{\bar q} = \alpha_Y^q,  \quad \quad \alpha_Q^{\bar q} = -\alpha_Q^q  \label{f:charge}
\end{align}
under the charge conjugation transformation and
\begin{align}
 & \alpha_Y^{u} = \alpha_Y^d,  \quad \quad \alpha_Q^{u} = -\alpha_Q^d  \label{f:symmetry}
\end{align}
under the $SU(2)$ symmetry (isospin).  In this paper, we only consider the $Q$-criterion.

In addition to the handedness, it is more convenient to introduce the handedness correlation in the electron positron annihilation process. On the one hand, if one does not distinguish the quark and antiquark jets, the total jet handedness of the $Y$-criterion vanishes since $H_Y^{\bar q}=-H_Y^{q}$ \cite{Efremov:1995fe}. On the other hand, the rather small value of the jet handedness regarded as cancellations of different flavor terms was difficult to measure.

To define the handedness correlation,  we first of all introduce the following probabilities,
\begin{align}
& \frac{N_R^+}{n^+}=\frac{N_L^-}{n^-}=\frac{1}{2}(1+\alpha), \label{f:alphapos} \\
& \frac{N_L^+}{n^+}=\frac{N_R^-}{n^-}=\frac{1}{2}(1-\alpha),  \label{f:alphanag}
\end{align}
where $n^{+,-}$ denote the numbers of the right-handed (left-handed) quarks, $N_{R,L}^{+,-}$ denote the numbers of the right-handed (left-handed) jets fragmented from the right-handed (left-handed) quarks. From Eqs. (\ref{f:alphapos}) and (\ref{f:alphanag}), we can write $\alpha$ as,
\begin{align}
 \alpha=\frac{N_R^+-N_L^+}{n^+}=\frac{N_L^- - N_R^-}{n^-}. \label{f:alpha}
\end{align}
We can see that $\alpha$ is the difference of probabilities of a right-handed (left-handed) quark to fragment into right-handed (left-handed) jets and left-handed (right-handed) jets.
As a result of Eqs. (\ref{f:alphapos}) and (\ref{f:alphanag}), the handedness can be rewritten as
\begin{align}
 H= \frac{N_R-N_L}{N}=\alpha P, \label{f:handLT}
\end{align}
which is consistent with Eq. (\ref{f:longH}).
We note that $N=N_R+N_L=n^+ + n^-$, $N_R=N_R^+ +N_R^-$ and $N_L=N_L^+ +N_L^-$. $P$ is the quark (longitudinal) polarization and defined as $ P=(n^+ - n^-)/N$.

The opposite jets handedness correlation can be defined as the minus product of the quark jet handedness and the antiquark jet handedness,
\begin{align}
 C^{q,\bar q}=-H^qH^{\bar q}, \label{f:handcordef}
\end{align}
where superscript $q, \bar q$ denote the quark and the corresponding antiquark. Using Eq. (\ref{f:handLT}), we have
\begin{align}
 C^{q,\bar q}=\alpha^q\alpha^{\bar q}. \label{f:handcorij}
\end{align}
Here we have omitted the quark-antiquark polarization correlation $c^{q,\bar q}$. In electron positron annihilation process, $c^{q,\bar q}=1$. Thus, it can be reduced in the formula.

Alternatively, the opposite jet handedness correlation can also be defined as
\begin{align}
 C^{q,\bar q}=\frac{N_{RL}+N_{LR}-N_{RR}-N_{LL}}{N_{RL}+N_{LR}+N_{RR}+N_{LL}}. \label{f:corredef}
\end{align}
Since helicities of quark and antiquark are correlated in annihilation process ($c^{q,\bar q}=1$), i.e, $n_{q,\bar q}^{++}=n_{q,\bar q}^{--}=0$. According to definitions in Eqs. (\ref{f:alphapos}) and (\ref{f:alphanag}), we have
\begin{align}
 N_{RR} &= n^{+-}_{q\bar q}\frac{1}{4}(1+\alpha^q)(1-\alpha^{\bar q})\nonumber \\
             &+n^{-+}_{q\bar q}\frac{1}{4}(1-\alpha^q)(1+\alpha^{\bar q}),  \label{f:NRR}\\
 N_{LL} &= n^{-+}_{q\bar q}\frac{1}{4}(1+\alpha^q)(1-\alpha^{\bar q})\nonumber\\
             &+n^{+-}_{q\bar q}\frac{1}{4}(1-\alpha^q)(1+\alpha^{\bar q}), \\
N_{RL} &= n^{+-}_{q\bar q}\frac{1}{4}(1+\alpha^q)(1+\alpha^{\bar q})\nonumber \\
             &+n^{-+}_{q\bar q}\frac{1}{4}(1-\alpha^q)(1-\alpha^{\bar q}), \\
 N_{LR} &= n^{-+}_{q\bar q}\frac{1}{4}(1+\alpha^q)(1+\alpha^{\bar q})\nonumber\\
             &+n^{+-}_{q\bar q}\frac{1}{4}(1-\alpha^q)(1-\alpha^{\bar q}). \label{f:NLR}
\end{align}
Substituting Eqs. (\ref{f:NRR})-(\ref{f:NLR}) into  Eq. (\ref{f:corredef}), one can obtain Eq. (\ref{f:handcorij}). From Eq. (\ref{f:charge})  the handedness correlation for opposite jets with $Q$-criterion can be rewritten as
\begin{align}
C_{Q}^{q,\bar q}=-\left(\alpha_{Q}^q \right)^2. \label{f:handcoQo}
\end{align}
It is not difficult to calculate the handedness correlation in the same jet with the same method. We have
\begin{align}
C_{Q}^{q, q}=\left(\alpha_{Q}^q \right)^2. \label{f:handcoQs}
\end{align}
%We can see that the opposite jets handedness correlation gives minus sign while same jet handedness correlation gives plus sign.
If one sum over the quark flavors, one can obtain
\begin{align}
 & C_{Q}^{q,\bar q}=-\frac{\sum_q\sigma_q \omega_q^2 (\alpha_Q^q)^2}{\sum_q\sigma_q \omega_q^2}, \\
 & C_{Q}^{q, q}=\frac{\sum_q\sigma_q \omega_q^2 (\alpha_Q^q)^2}{\sum_q\sigma_q \omega_q^2},
\end{align}
where $\sigma_q$ is the cross section of the flavor $q$ and $\omega_q$ is a probability of the flavor to fragment into the certain pair hadron.

One can see that theoretical calculation predicts negative sign for the opposite jet handedness correlation and positive sign for the same jet handedness correlation. This contradicts to experimental results. Based on this fact,  a straightforward consideration is how the positive quark polarization correlation ($c^{q,\bar q}=1$) transfers to the negative jet handedness correlation ($C_{Q}^{q,\bar q}=-\left(\alpha^q \right)^2$). What mechanism dominates this process? In addition to the valence quark polarization, there must be other origins of the handedness correlation, e.g, chromo-magnetic component of the gluon condensate of QCD vacuum \cite{Efremov:1995ff}.
In the following sections, we will propose a chromo-hydrogen-like model to explain this contradiction.  There is no need to introduce the QCD CP-violating effect in this model.

\section{The chromo-hydrogen-like model} \label{sec:model}

The chromo-hydrogen-like model in this section is proposed to describe the first stage of the fragmentation process rather than full stage. It is named after the mechanism of the hyperfine structure of hydrogen atoms. Particles in this model are combined according to their spin orientations. As a result, different ground states emerge.

In order to illustrate this model, we consider a valence quark $Q$ produced in the electron positron annihilation reaction with definite helicity. To be explicit, we use $S_Q$ and $Y_Q$ to denote its spin and color charge. In this case, we can write down the chromo-magnetic moment as we do in electrodynamics,
\begin{align}
 \vec \mu_Q=\frac{Y_Q}{2m_Q} \vec S_Q, \label{f:muQ}
\end{align}
where $m_Q$ is the mass of the valence quark in this model. The $g$-factor is not considered here. The moment shown above can induce a chromo-magnetic field \cite{Griffiths:2005},
\begin{align}
 \vec B_Q=\frac{1}{4\pi r^3}\left[3(\vec \mu_Q\cdot \hat r)\hat r - \vec\mu_Q\right] +\frac{2}{3}\vec \mu_Q \delta^3(\vec r), \label{f:magneticfield}
\end{align}
where $\hat r$ is the unit vector in the radial direction, $\hat r=\vec r/r$, $r$ is the distance between field point and the moment. The first term in Eq. (\ref{f:magneticfield}) is obtained by calculating the curl of the potential
\begin{align}
\vec A =\frac{1}{4\pi}\frac{\vec \mu_Q \times \hat r}{r^2}. \label{f:potential}
\end{align}
The second term in Eq. (\ref{f:magneticfield}) is introduced to describe the field at $r=0$.

To illustrate the chromo-hydrogen-like model, we consider the first stage of the fragmentation process, a quark-antiquark pair ($q,\bar q$) is produced in the chromo-magnetic field which is induced by the valence quark $Q$. Without loss of generality, we consider the interaction between $Q$ and $q$. QCD is not sensitive to the electric charge, ($Q,\bar q$) would give the same result. The chromo-magnetic moment of sea quark $q$ is
\begin{align}
 \vec \mu_q=\frac{Y_q}{2m_q} \vec S_q, \label{f:muq}
\end{align}
where $Y_q, m_q$ and $\vec S_q$ are the color charge, mass and spin vector of the quark $q$. Therefore, the Hamiltonian of the quark $q$ in the chromo-magnetic field due to the $Q$ magnetic moment is given by
\begin{align}
 \hat H =& -\mu_q \cdot \vec B_Q \nonumber\\
 =&-\frac{1}{4\pi}\frac{1}{r^3}\left[3(\vec \mu_Q\cdot \hat r)(\vec \mu_q \cdot \hat r) - \vec\mu_Q\cdot \vec\mu_q\right]  \nonumber\\
 &-\frac{2}{3}\vec \mu_Q\cdot\vec\mu_q \delta^3(\vec r). \label{f:hamiltonian}
\end{align}
Here we use $\hat H$ do denote the Hamiltonian in order to distinguish it from the jet handedness. Equation (\ref{f:hamiltonian}) is symmetric in $\vec \mu_Q$ and $\vec \mu_q$, as it should be.

If quarks $q$ and $Q$ form a hydrogen-like system, the Hamiltonian shown in Eq. (\ref{f:hamiltonian}) can be taken as a small perturbation. The first order correction to the system energy can be written as
\begin{align}
 E'=\langle \psi_0|\hat H| \psi_0\rangle, \label{f:perturbation}
\end{align}
where $|\psi_0 \rangle$ is the ground state wave function of this system which can be borrowed from the one for the atomic hydrogen ($S$-wave),
\begin{align}
 |\psi_0 \rangle =\frac{1}{\sqrt{\pi a^3}}e^{-r/a}  |\tilde S \rangle. \label{f:psi0}
\end{align}
Here $a$ is the Bohr radius and $|\tilde S \rangle$ is the spin state. Substituting Eqs. (\ref{f:hamiltonian}) and (\ref{f:psi0}) into Eq. (\ref{f:perturbation}) yields
\begin{align}
 E' &=-\frac{2}{3}\frac{1}{\pi a^3}\langle \vec \mu_Q\cdot\vec\mu_q \rangle \nonumber\\
  &=-\frac{2}{3}\frac{1}{\pi a^3}\frac{Y_Q Y_q}{4m_Q m_q}\langle \vec S_Q\cdot\vec S_q \rangle. \label{f:energy}
\end{align}

The vanishing of the contribution of the first term in Eq. (\ref{f:hamiltonian}) can also be understood in the following way. Replacing moments by spins and ignoring coefficients, we define
\begin{align}
 V=\frac{3(\vec S_Q\cdot \vec r)(\vec S_Q\cdot \vec r)}{r^2}-\vec S_Q \cdot \vec S_q. \label{f:vs}
\end{align}
Using $\vec S=\vec S_Q+\vec S_q$ and spin $\vec s=\vec \sigma/2$ ($\sigma$ is the Pauli matrix), we rewrite $V$ as
\begin{align}
 V=\frac{3(\vec S\cdot \vec r)^2}{2r^2}-\frac{S^2}{2}. \label{f:vss}
\end{align}
Since $\vec S\cdot \vec r= Sr\cos\theta$ with $\theta$ being the angle between $\vec S$ and $\vec r$, we have
\begin{align}
 V=S^2 \left(\frac{3}{2}\cos^2\theta-\frac{1}{2}\right)=S^2P_2, \label{f:vsw}
\end{align}
where $P_2$ is the Legendre function for $l=2$ which corresponds to the angular part of the $D$-wave. Therefore, the contribution of this term must vanish in the ground state ($S$-wave).

It is known that two spin-1/2 quarks can form either a triplet state or a singlet state according to spin combinations. Since
\begin{align}
 \vec S_Q\cdot \vec S_q =\frac{1}{2}\left[ S^2- S_Q^2 - S_q^2\right],
\end{align}
therefore, we have
\begin{align}
 E'_1= - \frac{1}{6}\frac{1}{\pi a^3}\frac{Y_Q Y_q}{4m_Q m_q}
\end{align}
for the triplet state and
\begin{align}
 E'_0= \frac{1}{2}\frac{1}{\pi a^3}\frac{Y_Q Y_q}{4m_Q m_q}
\end{align}
for the singlet state. If  the chromo-hydrogen-like state is colorless ($Q$ and $q$ have opposite color charges, $Y_Q Y_q<0$),  the singlet state would have lower energy level. In other words, the quark spins are antiparallel, or the chromo-magnetic moments are parallel. On the contrary, the triplet state would have lower energy level, i.e, the quark spins are parallel, or the chromo-magnetic moments are antiparallel.  Either way, the coupled sea quark will be polarized because of the certain polarized valence quark.

According to the calculation, we find that valence quark $Q$ and sea quark $q$ can form a chromo-hydrogen-like system which is non-degenerate. Different states lead to different fragmentation probabilities. Under this circumstance, subsequent fragmentation processes or hadron pair production process can be distinguished according to the combined states (triplet and/or singlet states) with different production probabilities, at least in principle. In the following section, we will recalculate the jet handedness and handedness correlation based on this interpretation and find that they have more than one origin.

\section{Handedness and handedness correlation. II} \label{sec:handednessrevisit}

In the previous section we presented the chromo-hydrogen-like model and illustrated quarks would combine together due to the color-induced interactions at the first stage of the fragmentation process. The non-degenerate states in the chromo-hydrogen-like model naturally divides the fragmentation process into two parts. In order to calculate the jet handedness we assume in this section that the triplet and singlet states fragment into either right-handed or left-handed jets independently.

According to Eqs. (\ref{f:alphapos})-(\ref{f:alphanag}), we can similarly write down the following equations,
\begin{align}
 & \frac{N_R^{+1}}{n^{+1}}=\frac{N_L^{-1}}{n^{-1}}=\frac{1}{2}(1+\alpha_h), \label{f:alphaposone}\\
 & \frac{N_L^{+1}}{n^{+1}}=\frac{N_R^{-1}}{n^{-1}}=\frac{1}{2}(1-\alpha_h), \label{f:alphanegone}\\
 & \frac{N_R^{+0}}{n^{+0}}=\frac{N_L^{-0}}{n^{-0}}=\frac{1}{2}(1+\beta_h), \label{f:betaposone}\\
 & \frac{N_L^{+0}}{n^{+0}}=\frac{N_R^{-0}}{n^{-0}}=\frac{1}{2}(1-\beta_h), \label{f:betanegone}
\end{align}
where $n^{\pm}=n^{\pm 1}+n^{\pm 0}$ and indices $1, 0$ denote triplet and singlet, respectively. We also use subscript $h$ to denote parameters in the chromo-hydrogen-like model. $N$ and $n$ with indices are also used to denote  numbers of jet and quark in Eqs (\ref{f:alphaposone})-(\ref{f:betanegone}).
The parameters $\alpha_h$ and $\beta_h$ are given by
\begin{align}
 &\alpha_h=\frac{N_R^{+1}-N_L^{+1}}{n^{+1}}=\frac{N_L^{-1}-N_R^{-1}}{n^{-1}},  \label{f:alphaequ} \\
 &\beta_h=\frac{N_R^{+0}-N_L^{+0}}{n^{+0}}=\frac{N_L^{-0}-N_R^{-0}}{n^{-0}}. \label{f:betaequ}
\end{align}
With conventions used above we can write down polarizations of the quarks $Q$ and $q$, respectively,
\begin{align}
 &P_Q=\frac{n^{+1}+n^{+0}-n^{-1}-n^{-0}}{N}, \label{f:PQpol}\\
 &P_q=\frac{n^{+1}+n^{-0}-n^{-1}-n^{+0}}{N}, \label{f:Pqpol}
\end{align}
where $N=n^{+1}+n^{+0}+n^{-1}+n^{-0}$.

For the jet handedness, as mentioned before, it is defined as the probability difference of the right-handed and left-handed jet, see Eq. (\ref{f:handLT}). In this section we redefine the jet handedness in a similar way by calculating both contributions from triplet and singlet states. Considering the probability difference of the right-handed and left-handed jets, we have
\begin{align}
 H_h &=\frac{N_R-N_L}{N}=\frac{(n^{+1}-n^{-1})\alpha_h+(n^{+0}-n^{-0})\beta_h}{N} \nonumber\\
 &=\frac{1}{2}(\alpha_h+\beta_h)P_Q + \frac{1}{2}(\alpha_h-\beta_h)P_q. \label{f:handednessHM}
\end{align}
In order to obtain this equation, we have used Eqs. (\ref{f:alphaposone})-(\ref{f:betanegone}) and Eqs. (\ref{f:PQpol})-(\ref{f:Pqpol}).
Equation (\ref{f:handednessHM}) implies the origin of the jet handedness can be divided into two parts, one is proportional to to the polarization of the valence quark $Q$ and the other is proportional to the polarization of the sea quark $q$. This result is different from the one in Eq. (\ref{f:handLT}) in which jet handedness is only a function of the polarization of quark $Q$ ($P_Q$).

According to Eq. (\ref{f:handcordef})  we can  immediately write down the jet handedness correlation,
\begin{align}
 C_h^{q,\bar q}=&\frac{1}{4}\bigg[(\alpha_h +\beta_h)^2P_Q^q P_Q^{\bar q}+(\alpha_h -\beta_h)^2P_q^q P_q^{\bar q}\nonumber\\
& +(\alpha_h^2 -\beta_h^2)\big(P_Q^{q} P_q^{\bar q}+P_q^{q} P_Q^{\bar q}\big)\bigg] \label{f:correlationHM}
\end{align}
for the opposite jets correlation.  According to the method illustrated in Eqs. (\ref{f:corredef})-(\ref{f:NLR}), one can obtain the same result with the definition in Eq. (\ref{f:corredef}),
\begin{align}
 C_h^{q,\bar q}=-\frac{\left[(n^{+1}-n^{-1})\alpha_h+(n^{+0}-n^{-0})\beta_h\right]^2}{(n^{+1}+n^{+0}+n^{-1}+n^{-0})^2}.
\end{align}
For the same jet correlation, we have $ C_h^{q, q}=- C_h^{q,\bar q}$. The first term in Eq. (\ref{f:correlationHM}) corresponds to Eq. (\ref{f:handcoQo}) when set $P_Q^q P_Q^{\bar q}=-1$. The second and third terms give the alternative contributions to the handedness correlation. They dominate the sign of the handedness correlation in the chromo-hydrogen-like model.

Form Eq. (\ref{f:correlationHM}), we see that $C^{q, \bar q}$ can be positive definite as long as the sea quark $q$ polarization is taken into consideration. This conclusion is consistent with experimental results. Let's explain this. After being produced from the annihilation reaction, valence quark $Q$ would induce a chromo-magnetic filed due to its chromo-magnetic moment. At the first stage of the fragmentation process, sea quark pair ($q, \bar q$) is generated in the color field. Soon afterwards, sea quarks combine with $Q$ and form chromo-hydrogen-like systems. As a result, sea quark is polarized due the polarization of $Q$, see Eq. (\ref{f:energy}).
%($P_Q + P_q =2(n^{+1}-n^{-1})/N, P_Q - P_q =2(n^{+0}-n^{-0})/N$).
Both polarizations of $Q$ and $q$ can be transferred to the final jets (jet handedness, Eq. (\ref{f:correlationHM})) and hadrons (hadron polarization) via triplet and singlet spin combinations in the fragmentation process.

\begin{figure}[t]
  \centering
 \includegraphics[width=5cm]{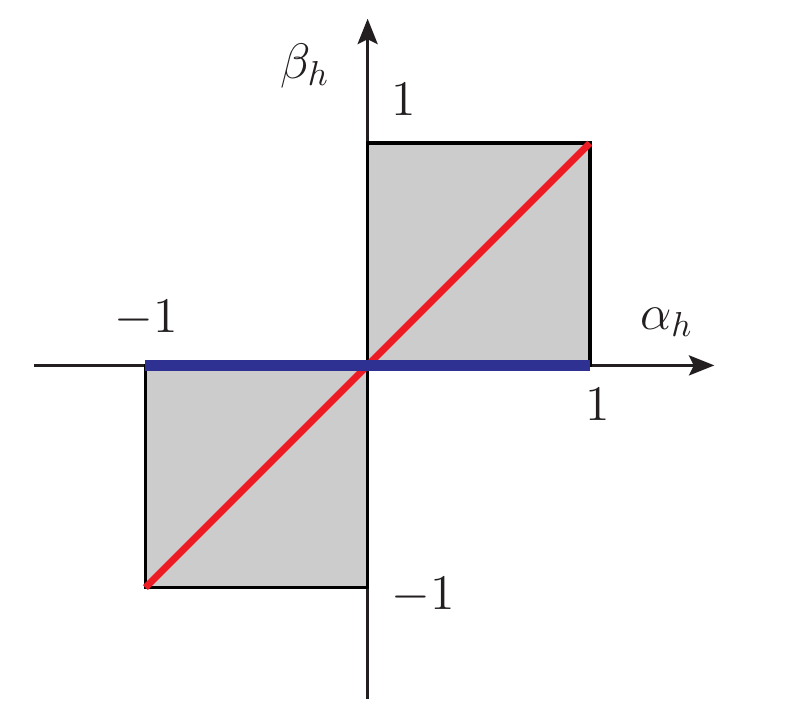}\\
  \caption{The values region of parameters of  $\alpha, \beta$.}\label{fig:handj}
\end{figure}

To end this section, we present a discussion of the jet handedness and the handedness correlation with respect to parameters $\alpha_h$ and $ \beta_h$.  We define a parameter plane and show the values region in Fig. \ref{fig:handj}.
\begin{itemize}
\item In order to calculate the jet handedness in the chromo-hydrogen-like model we assume that the triplet and singlet states fragment into either right-handed or left-handed jets independently. According to Eqs. (\ref{f:alphaequ})-(\ref{f:betaequ}), we can see that $\alpha_h$ and $\beta_h$ have same signs because of the same definitions and interpretations.  They are illustrated in the shadow area in Fig. \ref{fig:handj}. As a result, we have the following constraint,
\begin{align}
 |\alpha_h-\beta_h|<|\alpha_h+\beta_h|.
\end{align}
Since sea quark polarization $P_q$ originates from the valence quark polarization $P_Q$, $P_q$ should be smaller than $P_Q$, i.e, $P_q<P_Q$. Thus the valence quark polarization term would dominate the magnitude value of the handedness. This can also be understood in terms of higher order perturbative process.

\item If we require $\alpha_h = \beta_h$, Eqs. (\ref{f:handednessHM}) and (\ref{f:correlationHM}) will reduce to Eqs. (\ref{f:handLT}) and (\ref{f:handcoQo}), respectively. The identity implies there is no difference between the triplet and singlet states, or no non-degenerate states are formed, or no chromo-hydrogen-like systems are formed. In this case, sea quark $q$ polarization $P_q$ vanishes and only the quark $Q$ polarization $ P_Q$ emerges and dominates the fragmentation process. Parameter points will be on the red line in Fig. \ref{fig:handj}. If possible, we name this line as Q-line which can be used to determine the quark $Q$ polarization. In this case, our discussion will be reduced to the original one.

\item If $\beta_h=0$, $P_Q$ and $P_q$ have same contributions to the jet handedness and handedness correlation,
\begin{align}
 &H_h=\frac{1}{2}\alpha_h \left(P_Q+P_q\right), \label{f:Hb0}\\
 &C^{q, \bar q}_h=\frac{1}{4}\alpha_h^2\left(P_Q^qP_Q^{\bar q}+P_q^qP_q^{\bar q} + P_Q^{\{ q}P_q^{\bar q\}}\right). \label{f:Cb0}
\end{align}
where $\{A, B\}= AB+BA$. The jet handedness and handedness are symmetric in $P_Q$ and $P_q$. This is because right-handed jets and left-handed jets have the same probabilities to be fragmented from the combined states.
In this case, the value of $\alpha_h$ will be limited to the $\alpha_h$-axis (the blue line in Fig. \ref{fig:handj}),  as it should be.  Mathematically, if one  requires $\alpha_h=0$,  one can obtain
\begin{align}
 &H_h=\frac{1}{2}\beta_h \left(P_Q-P_q\right), \label{f:Ha0}\\
 &C^{q, \bar q}_h=\frac{1}{4}\beta_h^2\left(P_Q^qP_Q^{\bar q}+P_q^qP_q^{\bar q} - P_Q^{\{ q}P_q^{\bar q\}}\right). \label{f:Ca0}
\end{align}
%The jet handedness is given by the difference of $P_Q$ and $P_q$.
Here we have
\begin{align}
 & P_Q + P_q =\frac{2(n^{+1}-n^{-1})}{N}, \\
 & P_Q - P_q =\frac{2(n^{+0}-n^{-0})}{N}.
\end{align}
\end{itemize}

From the previous arguments, we know both the valence quark polarization  $P_Q$ and the sea quark polarization $P_q$ contribute to the jet handedness and handedness correlation. 

In addition to explaining the contradiction between experimental results and theoretical predictions, the chromo-hydrogen-like model gives a method to determine the sea quark polarization by measuring jet handedness and/or handedness correlation as long as the valence quark polarization is known \cite{Chen:2016iey}.
So the question that remains is how to determine the values of $\alpha_h$ and $ \beta_h$. Theoretically, we do not know the numbers of triplet and singlet states and the determination of $\alpha_h$ and $ \beta_h$ is impossible. Maybe phenomenological model can do this, but it is still an open question.

\section{Summary} \label{sec:summary}

Jet handedness used to be taken as a tool to determine quark and/or gluon polarizations. It was demonstrated by measuring the correlation between jet handedness in the electron positron annihilation process.  Once parameters are determined, the method could be used to measure quark and/or gluon polarizations in other experiments. Although, reported data provided evidence for the jet handedness and handedness correlation, there is a contradiction between the experimental results and theoretical predictions for the opposite jets handedness correlation. In order to solve this problem, we recalculate the jet handedness and the handedness correlation in the electron positron annihilation process based on the chromo-hydrogen-like model.
In this model, the valence quark $Q$ produced in the annihilation reaction and a sea quark $q$ form a chromo-hydrogen-like system due to interactions between chromo-magnetic moments or interactions between spins. The ground state of this system is non-degenerate and splits into a triplet state and a singlet state. Assuming the triplet state and the singlet state fragment independently into right-handed and/or left-handed jets, we rewrite down the jet handedness  and the handedness correlation in terms of parameters  $\alpha_h$ and $ \beta_h$ which are production probability differences of the right-handed and left-handed jets for the triplet and singlet states, respectively.
According to our calculation, we find that both valence quark polarization ($P_Q$) and the sea quark polarization  ($P_q$)  contribute to the jet handedness and handedness correlation, but $P_q<P_Q$. Parameters  $\alpha_h$ and $ \beta_h$ satisfy $|\alpha_h-\beta_h|<|\alpha_h+\beta_h|$.
If $\alpha_h=\beta_h$, the sea quark polarization $P_q$ vanishes and the valence quark polarization $ P_Q$ dominates the fragmentation process.  If $\beta_h=0$, $P_Q$ and $P_q$ have same contributions to the jet handedness and handedness correlation. This is because right-handed jets and left-handed jets have the same probabilities to be fragmented from the combined states.  Apart from explaining the contradiction, our calculation gives a method to determine the sea quark polarization by measuring jet handedness and/or handedness correlation as long as $\alpha_h$ and $ \beta_h$ are known.

\section*{Acknowledgements}

This work was supported by the Natural Science Foundation of Shandong Province (Grant No. ZR2021QA015).

%\appendix

%\section{}

\end{CJK*}
\end{document}